\documentclass[pdflatex,sn-mathphys-num]{sn-jnl}

\usepackage{amsmath,amssymb}
\usepackage{xcolor}
\usepackage{booktabs}
\usepackage{graphicx}
\usepackage{array}
\usepackage[title]{appendix}

\hypersetup{hidelinks,
  pdftitle={Shortening agent skills leaves LLM agent cost unchanged},
  pdfauthor={Xiaonan Xu and Wenjing Wu}
}

\def\burl#1{\url{#1}}

\newcommand{\minus}{\ensuremath{-}}

\begin{document}

\title[Shortening agent skills leaves LLM agent cost unchanged]{Shortening agent skills leaves LLM agent cost unchanged}

\author*[1]{\fnm{Xiaonan} \sur{Xu}}\email{xiaonanxu5@gmail.com}

\author[2]{\fnm{Wenjing} \sur{Wu}}\email{wuwenjing256@gmail.com}

\affil*[1]{\orgdiv{Computer Information Technology}, \orgname{Northern Arizona University}, \orgaddress{\city{Flagstaff}, \state{AZ}, \country{USA}}}

\affil[2]{\orgdiv{Computer Science}, \orgname{University of Colorado Boulder}, \orgaddress{\city{Boulder}, \state{CO}, \country{USA}}}

\abstract{Agent skills are often shortened, by compression, rewriting with a language model or partial loading, to cut the cost of running LLM agents. We tested whether this works on 40 SkillsBench tasks. The agent received a curated skill either as written or in one of several shortened versions, each version ran on a small and a large model three times per task, and every run was priced at the provider's published rates. The shortened skills were up to 54\% shorter, and cost per task stayed at or slightly above that of the original skill, because the skill is a small and mostly cached part of the text the agent processes. Shortening by a program that keeps the original sentences kept the success rate at the level of the original skill. Rewriting by a language model lowered success on the small model, from 42.5\% with the original skill to between 29.2\% and 37.5\%, and a stronger rewriting model gave the same result. The largest change came from the model that runs the agent: moving to the large model raised success by 18 to 27 points at 4.6 to 6.5 times the cost. The results favour using curated skills as written and spending effort on the choice of agent model.}


\maketitle

\section{Introduction}\label{sec:intro}

Agent skills give a large language model (LLM) agent procedural knowledge for a type of task. A skill is a folder centred on a SKILL.md file, with optional scripts and reference files, and the agent loads it when it works on a matching task \cite{zhang2025skills}. On SkillsBench, curated skills written by people raised agent success rates across many combinations of model and agent harness, while skills that the models wrote for themselves gave no average benefit \cite{li2026skillsbench}.

A skill adds text to the agent's context, and several lines of work shorten skills to reduce the cost of running agents. Compression methods remove redundant text \cite{gao2026skillreducer,bai2026skillzip}, structured representations turn a skill into a compact schema \cite{liang2026ssl}, and progressive disclosure loads only the part of a skill that a task needs \cite{zhang2025skills}. These methods report shorter skills with similar success rates, and most of them measure the saving in tokens.

Two things make the saving uncertain. An agent run is a long sequence of model calls in which the agent reads files, runs commands and reads their output, and providers bill input that repeats across calls at a lower cached rate. The skill is one part of this text, so a shorter skill may change the bill very little. Most ways of shortening a skill also rewrite it, and a rewrite can lose instructions that the agent needs. Xing et al.\ found that shorter skills can increase the tokens an agent uses afterwards, because removed details lead the agent to explore more \cite{xing2026skillremember}, and Dong et al.\ traced skill cost to the extra steps a skill induces as well as to its length \cite{dong2026harmful}. Our earlier study on SkillsBench found that changing how finely a curated skill is presented had small effects that varied by model \cite{xu2026granularity}. The present study moves from presentation to shortening and measures the provider bill directly.

We ran 40 SkillsBench tasks with the original skill and with shortened versions produced in different ways: a program that keeps a subset of the original sentences, a language model that rewrites the skill as plain text or as a structured document, and partial loading. Each version ran on a small and a large model, and every run was priced at the provider's published rates with caching as recorded. We address three questions. RQ1 asks whether shortening a skill lowers the cost of running the agent. RQ2 asks how shortening affects the success rate, and whether it matters if a program or a language model does the shortening. RQ3 asks which has the larger effect, the model that rewrites the skill or the model that runs the agent.

Shortening kept cost at or slightly above the level of the original skill. A program that keeps the original sentences kept the success rate, while rewriting by a language model lowered it on the small model, whichever model did the rewriting. The model that runs the agent was the only factor whose effect held after correction for multiple comparisons. Section~\ref{sec:related} reviews related work, Section~\ref{sec:design} describes the study, Section~\ref{sec:results} reports the results, Section~\ref{sec:discussion} discusses why they arise and what developers can do, Section~\ref{sec:scope} states the scope of the data and Section~\ref{sec:conclusion} concludes.

\section{Related work}\label{sec:related}

\textbf{Usefulness of skills.} SkillsBench introduced paired evaluation of agents with and without curated skills on tasks with automatic verifiers, and found that skills generated by the agent's own model give little benefit \cite{li2026skillsbench}. SkillAxe refines skills written by models with feedback from agent runs and closes part of the gap to skills written by people \cite{gautam2026skillaxe}. We use SkillsBench tasks and their curated skills. Our model rewrites start from a curated skill and restate it, which differs from writing a skill from scratch.

\textbf{Shortening and rewriting skills.} SkillReducer shortens skill descriptions and bodies and loads detail only when needed, and reports that task quality is preserved \cite{gao2026skillreducer}. SkillZip compresses skills by finding structure that repeats across them, without running evaluations \cite{bai2026skillzip}. Xing et al.\ treat rewriting as a choice of what a skill must keep, measure the agent's downstream tokens alongside the skill's own tokens, and learn to select a rewrite strategy \cite{xing2026skillremember}. Our program shortening and model rewrites are single-pass versions of these operations, compared against the original skill with cost taken from the provider bill.

\textbf{Presentation of skills.} Studies of prompt format show that the same content in different formats changes model performance, and that structured formats are not always better than plain text \cite{he2024promptformat}. The Scheduling-Structural-Logical representation converts SKILL.md files into a three-layer structured form \cite{liang2026ssl}. A study of small and medium open models in industrial settings reports that small models often miss hierarchical structure inside a SKILL.md file \cite{xu2026agentskillframework}. Our earlier SkillsBench study varied how finely curated skills are presented and found small, model-dependent effects \cite{xu2026granularity}. SkillJuror compares differently organised versions of the same task skill, with task scope and helper files held fixed, and measures how they change agent behaviour at run time \cite{skilljuror2026}. Here we hold the instruction content fixed and vary only the format, structured against plain text.

\textbf{Cost and the agent's model.} Dong et al.\ found that skills can induce unnecessary work and attributed skill cost to both context and the steps of the run \cite{dong2026harmful}. SKILLER optimises skills for a small model with language-level reinforcement learning and reports that a small model with an optimised skill can outperform a larger model with skills written by people \cite{dang2026skiller}. Our rewrites use no feedback from task runs, and Section~\ref{sec:executor} compares the two settings.

\section{Study design}\label{sec:design}

\subsection{Tasks}\label{sec:tasks}

We used SkillsBench \cite{li2026skillsbench}, in which each task provides a containerised environment, a task description, a curated skill and an automatic verifier that returns pass or fail. The curated skills are written by the benchmark contributors and filtered for quality, so every shortened version in this study starts from a skill that has already been curated. We selected 40 tasks that cover all eight benchmark domains (cybersecurity; finance and economics; industrial and physical systems; mathematics and formal reasoning; media and content production; natural science; office and white-collar work; software engineering) and all three difficulty levels, leaving out tasks that need a GPU or exceed local resource limits. Appendix~\ref{app:tasks} lists the tasks.

\subsection{Models}\label{sec:models}

The small model was gpt-5.4-mini and the large model was gpt-5.5, both used through dated snapshots. All runs used the benchmark's agent harness with one container per run, and each task was run three times for every condition.

\subsection{Skill versions}\label{sec:conditions}

Table~\ref{tab:conditions} lists the ten conditions. Besides the original skill and a no-skill baseline, there are four ways of shortening the skill. A program keeps a subset of the original sentences and makes no model call. For the two model rewrites, a language model first extracts a list of short instruction units from the original skill, and a fixed script then writes the same units either as plain running text or as headed sections in which each unit carries kind, priority and operation fields. The two rewrites contain identical units, so a difference between them comes from the format alone. Partial loading gives the agent a bounded selection of skill material chosen by a fixed selector. The rewriting model received the original skill, a fixed instruction and output format and a token budget, and it saw no task prompt, solution, verifier code or run trajectory. The structured rewrite was written by both models and run on both models, so the effect of the rewriting model can be separated from the effect of the model that runs the agent. Unless a condition name says otherwise, shortened skills were written by the large model and run on the small model.

\begin{table}[htbp]
\caption{Conditions}\label{tab:conditions}
\small
\begin{tabular}{@{}>{\raggedright\arraybackslash}p{3.9cm}>{\raggedright\arraybackslash}p{3.9cm}>{\raggedright\arraybackslash}p{1.6cm}>{\raggedright\arraybackslash}p{1.6cm}@{}}
\toprule
Condition & What the agent receives & Prepared by & Agent model \\
\midrule
No skill & no skill & -- & small \\
Original & the curated skill unchanged & -- & small \\
Program-shortened & a subset of the original sentences & program & small \\
Plain-text rewrite & instruction units as running text & large model & small \\
Structured rewrite & instruction units as headed sections & large model & small \\
Structured rewrite, written by small model & instruction units as headed sections & small model & small \\
Partial loading & a bounded selection of skill material & large model & small \\
Original, large model & the curated skill unchanged & -- & large \\
Structured rewrite, large model & instruction units as headed sections & large model & large \\
Structured rewrite, written by small model, large model & instruction units as headed sections & small model & large \\
\botrule
\end{tabular}
\end{table}

\subsection{Success and cost}\label{sec:outcomes}

A run succeeds when the benchmark's verifier passes it, and a run that times out counts as a failure. The success rate of a condition averages the three runs of each task and then the 40 tasks. Cost is the solve-stage bill in US dollars, computed from the recorded uncached input, cached input and output tokens of each run at the provider's published on-demand prices, with reasoning tokens counted as output. We call this the actual cost. Two other views appear where they help: cost without the caching discount, which charges all input at the uncached price, and token-count cost, which charges both models the same price per token and corresponds to the way most prior work reports savings. Skill length is the size of the skill given to the agent, in tokens. The one-time cost of producing each rewrite is reported separately.

\subsection{Statistics}\label{sec:stats}

The task is the unit of analysis. For each comparison we took the difference between two conditions on each task and averaged it over the 40 tasks. We report 95\% intervals from 10,000 bootstrap resamples of tasks \cite{efron1993bootstrap} and two-sided p-values from a sign-flip permutation within tasks. The eight comparisons in Table~\ref{tab:contrasts} are corrected together with the Holm procedure \cite{holm1979simple}.

\section{Results}\label{sec:results}

\subsection{The curated skill raised the success rate}\label{sec:passrate}

\begin{figure}[htbp]
\centering
\includegraphics[width=\textwidth]{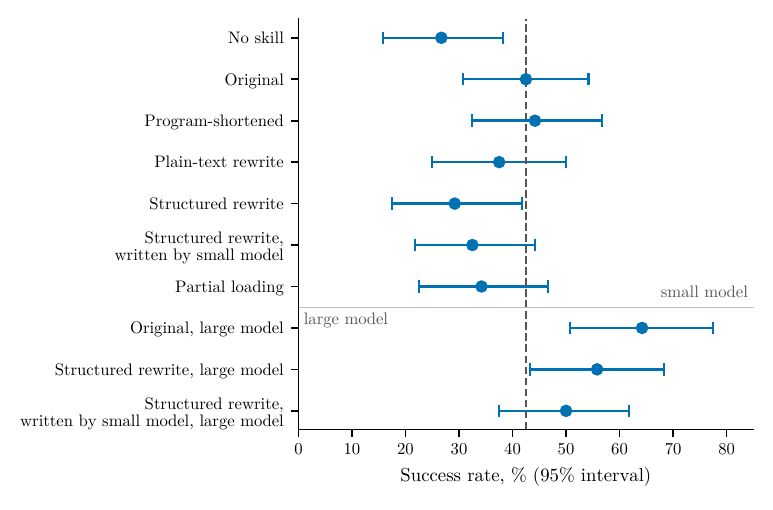}
\caption{Success rate by condition with 95\% intervals. The dashed line marks the original skill on the small model}\label{fig:passrate}
\end{figure}

Of the 1,200 runs, 500 passed, 647 failed and 53 timed out. Table~\ref{tab:passrate} and Figure~\ref{fig:passrate} give the success rate of each condition. On the small model the original skill raised the success rate from 26.7\% with no skill to 42.5\%, and on the large model it reached 64.2\%, the highest of all conditions.

\begin{table}[htbp]
\caption{Runs and success rate by condition (40 tasks, three runs each)}\label{tab:passrate}
\small
\begin{tabular}{@{}>{\raggedright\arraybackslash}p{4.6cm}rrrl@{}}
\toprule
Condition & Passed & Failed & Timed & Success rate, \% \\
 & & & out & [95\% interval] \\
\midrule
No skill & 32 & 85 & 3 & 26.7 [15.8, 38.3] \\
Original & 51 & 62 & 7 & 42.5 [30.8, 54.2] \\
Program-shortened & 53 & 60 & 7 & 44.2 [32.5, 56.7] \\
Plain-text rewrite & 45 & 71 & 4 & 37.5 [25.0, 50.0] \\
Structured rewrite & 35 & 79 & 6 & 29.2 [17.5, 41.7] \\
Structured rewrite, written by small model & 39 & 75 & 6 & 32.5 [21.7, 44.2] \\
Partial loading & 41 & 72 & 7 & 34.2 [22.5, 46.7] \\
Original, large model & 77 & 39 & 4 & 64.2 [50.8, 77.5] \\
Structured rewrite, large model & 67 & 49 & 4 & 55.8 [43.3, 68.3] \\
Structured rewrite, written by small model, large model & 60 & 55 & 5 & 50.0 [37.5, 61.7] \\
\botrule
\end{tabular}
\footnotetext{Intervals come from bootstrap resampling of tasks. Conditions on the small model are listed first.}
\end{table}

\subsection{Shortening left cost unchanged (RQ1)}\label{sec:cost}

Table~\ref{tab:cost} gives skill length, tokens and cost for each condition. The shortened skills ranged from 2,080 to 3,566 tokens, against 4,550 for the original. On the small model, total tokens per task stayed between 46,072 and 50,726 whatever the skill length, and actual cost stayed between \$0.0062 and \$0.0078 per task, with the original skill the cheapest. Each shortened version cost \$0.0004 to \$0.0013 more per task than the original. The program-shortened skill, the closest in cost, was cheaper than the original on 22 tasks and more expensive on 18. Across all runs, 95.0\% of input tokens were billed at the cached rate. The no-skill condition used about as many tokens as the original skill (47,435 against 47,529) and cost more (\$0.0078 against \$0.0062); without the caching discount the two cost the same (\$0.0374 against \$0.0370), so the gap comes from caching.

\begin{table}[htbp]
\caption{Skill length, tokens and cost per task by condition}\label{tab:cost}
\small
\begin{tabular}{@{}>{\raggedright\arraybackslash}p{3.7cm}rrrrr@{}}
\toprule
Condition & Skill & Total & Actual & Cost without & Token-count \\
 & length & tokens & cost, USD & caching, USD & cost, USD \\
\midrule
No skill & 0 & 47,435 & 0.0078 & 0.0374 & 0.0082 \\
Original & 4,550 & 47,529 & 0.0062 & 0.0370 & 0.0069 \\
Program-shortened & 3,566 & 46,089 & 0.0066 & 0.0363 & 0.0069 \\
Plain-text rewrite & 2,785 & 50,726 & 0.0066 & 0.0396 & 0.0074 \\
Structured rewrite & 2,981 & 48,981 & 0.0075 & 0.0383 & 0.0083 \\
Structured rewrite, written by small model & 2,525 & 49,695 & 0.0070 & 0.0387 & 0.0079 \\
Partial loading & 2,080 & 46,072 & 0.0071 & 0.0362 & 0.0078 \\
Original, large model & 4,550 & 41,532 & 0.0406 & 0.2125 & 0.0071 \\
Structured rewrite, large model & 2,981 & 42,183 & 0.0347 & 0.2158 & 0.0062 \\
Structured rewrite, written by small model, large model & 2,525 & 43,617 & 0.0400 & 0.2232 & 0.0068 \\
\botrule
\end{tabular}
\footnotetext{Skill length is in tokens. Actual cost uses published on-demand prices with caching as recorded. Token-count cost charges both models the same price per token.}
\end{table}

Producing the instruction units for the model rewrites cost \$5.40 for the 40 tasks, \$4.82 with the large model and \$0.58 with the small model. Every shortened version had an average cost at or above that of the original skill, so this one-time cost adds to the total.

\subsection{Model rewriting lowered the success rate (RQ2)}\label{sec:rq2}

\begin{figure}[htbp]
\centering
\includegraphics[width=\textwidth]{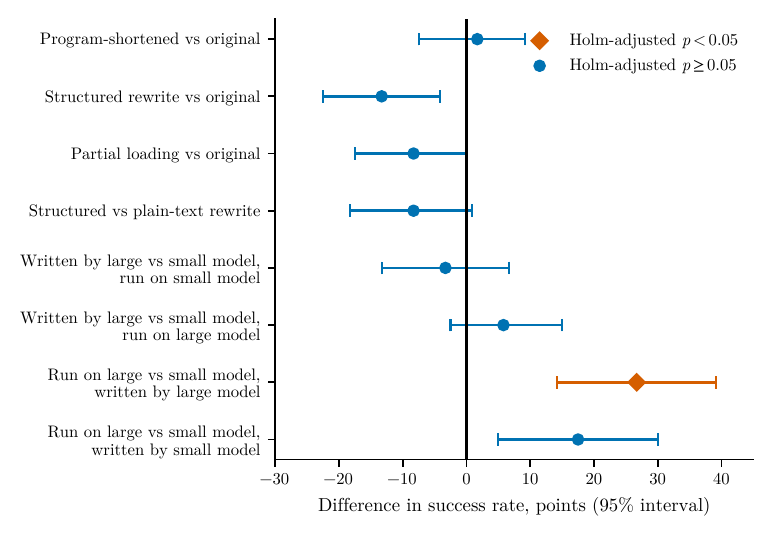}
\caption{Paired differences in success rate with 95\% intervals. The comparison that remains significant after Holm correction is drawn in vermilion with a diamond marker}\label{fig:contrasts}
\end{figure}

Table~\ref{tab:contrasts} and Figure~\ref{fig:contrasts} give the paired comparisons. The program-shortened skill kept the success rate at the original level, with a difference of +1.7 points [\minus7.5, +9.2]. The model rewrites scored lower. The plain-text rewrite reached 37.5\%, and the structured rewrite reached 29.2\%, 13.3 points below the original skill [\minus22.5, \minus4.2] with a Holm-adjusted p of 0.080. With identical instruction units, the structured format scored 8.3 points below plain text [\minus18.3, +0.8]. Partial loading scored 8.3 points below the original [\minus17.5, 0.0].

The model that did the rewriting made little difference. On the small model, the structured rewrite written by the large model scored 3.3 points below the one written by the small model [\minus13.3, +6.7], and on the large model it scored 5.8 points above [\minus2.5, +15.0].

\begin{table}[htbp]
\caption{Paired comparisons of success rate and actual cost (40 tasks)}\label{tab:contrasts}
\small
\begin{tabular}{@{}>{\raggedright\arraybackslash}p{4.8cm}llr@{}}
\toprule
Comparison & Difference in & Holm p & Difference in \\
 & success, points & & cost per task, USD \\
\midrule
Program-shortened vs original & +1.7 [\minus7.5, +9.2] & 1.000 & +0.00037 \\
Structured rewrite vs original & \minus13.3 [\minus22.5, \minus4.2] & 0.080 & +0.00131 \\
Partial loading vs original & \minus8.3 [\minus17.5, 0.0] & 0.553 & +0.00088 \\
Structured vs plain-text rewrite & \minus8.3 [\minus18.3, +0.8] & 0.591 & +0.00088 \\
Written by large vs small model, run on small model & \minus3.3 [\minus13.3, +6.7] & 1.000 & +0.00051 \\
Written by large vs small model, run on large model & +5.8 [\minus2.5, +15.0] & 0.872 & \minus0.00526 \\
Run on large vs small model, written by large model & +26.7 [+14.2, +39.2] & 0.003 & +0.02720 \\
Run on large vs small model, written by small model & +17.5 [+5.0, +30.0] & 0.080 & +0.03298 \\
\botrule
\end{tabular}
\footnotetext{The last four comparisons use the structured rewrite. Positive values favour the first condition named. Brackets give 95\% intervals.}
\end{table}

\subsection{The agent's model had the largest effect (RQ3)}\label{sec:rq3}

Changing the model that runs the agent had the largest effect. For the structured rewrite written by the large model, moving the agent from the small to the large model raised the success rate by 26.7 points [+14.2, +39.2], with a Holm-adjusted p of 0.003, the only comparison that held after correction. For the rewrite written by the small model, the same change raised it by 17.5 points [+5.0, +30.0], and for the original skill by 21.7 points. On the large model the agent cost \$0.035 to \$0.041 per task.

\begin{figure}[htbp]
\centering
\includegraphics[width=\textwidth]{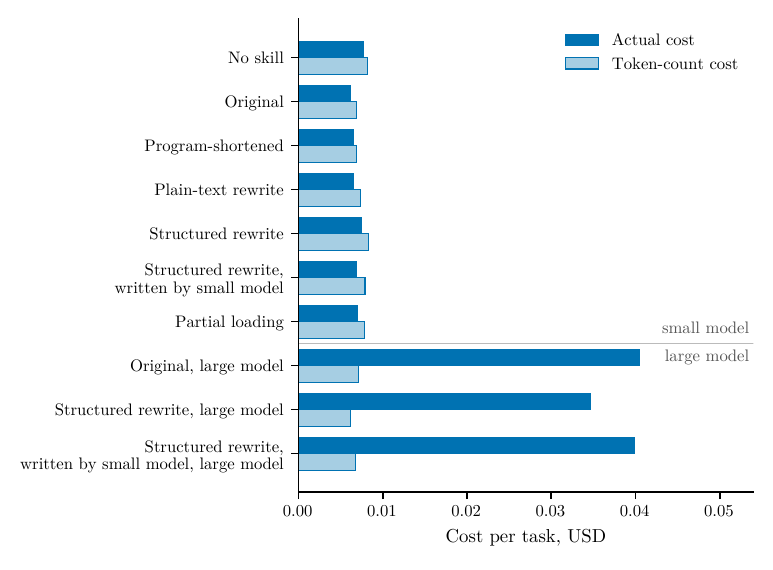}
\caption{Actual cost and token-count cost per task by condition}\label{fig:cost}
\end{figure}

Figure~\ref{fig:cost} shows how a token count would change this picture. The large model used fewer tokens per task (41,532 to 43,617), so with both models charged the same price per token it would cost \$0.0062 to \$0.0071, as little as the small-model conditions. The actual bill depends on the price of the model as much as on the number of tokens.

\section{Discussion}\label{sec:discussion}

\subsection{Why shortening left cost unchanged}\label{sec:why-cost}

Most of what an agent run costs comes from the agent's own work. On the small model, skill length varied by more than half across conditions while total tokens varied by about a tenth. The agent reads files, runs commands and reads their output over many calls, and each call sends the conversation so far. The skill sits near the start of this conversation, in the part that repeats from call to call and is billed at the cached rate, which made up 95\% of input in our runs. Removing part of the skill therefore takes tokens out of the cheapest part of the bill. This agrees with Xing et al., who found that the value of a rewrite depends on the agent's downstream tokens more than on the length of the skill \cite{xing2026skillremember}, and with Dong et al., who attributed skill cost to the steps a skill induces \cite{dong2026harmful}.

\subsection{Why model rewriting lowered the success rate}\label{sec:why-pass}

The program keeps the original sentences, and its version scored at the level of the original skill. Every model rewrite scored lower on the small model. When a language model restates a skill as short instruction units, it can drop or simplify details whose use only shows during the task. Xing et al.\ found that removing such details sends agents into longer exploration \cite{xing2026skillremember}, and SkillsBench found that skills written by models add little \cite{li2026skillsbench}. A larger rewriting model gave the same result, which indicates that the loss comes from restating the skill, whichever model does it. The structured format scored lower than plain text with the same units, in line with the report that small models often miss hierarchical structure in a SKILL.md file \cite{xu2026agentskillframework}, and on the large model the gap between the original skill and the structured rewrite was smaller, 8.4 points.

\subsection{The agent's model and trained skill optimisation}\label{sec:executor}

In our data the model that runs the agent had the largest effect. SKILLER reports a different pattern, in which a small model with an optimised skill outperformed a larger model \cite{dang2026skiller}. The two settings differ. SKILLER improves a skill using the outcomes of many task runs, while the shortening in this study is a single pass with no feedback from runs. A single-pass rewrite of a curated skill kept the small model at or below its success rate with the original skill, and the way to a higher success rate in our data was a larger model at a higher price. A change of model also changes which individual tasks succeed. Our earlier measurement of commercial LLM API upgrades found that upgrades with higher aggregate scores still contained items whose success reliably fell \cite{xu2026aggregate}, so a move to a larger model is worth checking task by task.

\subsection{What developers can do}\label{sec:practice}

Three practices follow from the measurements. Curated skills can be used as written: on the small model the original skill was the cheapest condition and within two points of the best success rate. When a shorter skill is needed, cutting it with a program that keeps the original sentences held the success rate at equal cost, while rewriting it with a language model lowered the success rate by 5.0 to 13.3 points. Savings are best compared on the actual bill with caching included, since the token count showed the large model as cheap as the small one while the bill showed it several times more expensive.

\section{Scope of the data}\label{sec:scope}

The data cover 40 SkillsBench tasks across eight domains and three difficulty levels, one small and one large model from a single provider, the benchmark's agent harness, three runs per task and condition, and the provider's on-demand prices at the time of the runs in 2026.

\section{Conclusion}\label{sec:conclusion}

Shortening a curated agent skill kept the cost of running an LLM agent at or slightly above the level of the original skill on 40 SkillsBench tasks, because the skill is a small and mostly cached part of what the agent processes. Shortening by a program kept the success rate, rewriting by a language model lowered it on the small model, and the model that runs the agent had the largest effect on both success and cost. For developers who want a cheaper or more reliable agent, the choice of model mattered more than the length of the skill.

\backmatter

\section*{Declarations}

\bmhead{Data availability}

Run outcomes, token counts, cost accounting, the skill variants and an analysis script are available at \url{https://doi.org/10.5281/zenodo.21148499}.

\bmhead{Author contributions}

Xiaonan Xu: conceptualisation, methodology, software, formal analysis, investigation, data curation, writing (original draft, review and editing). Wenjing Wu: methodology, validation, writing (review and editing).

\bmhead{Funding}

The authors received no external funding for this work.

\bmhead{Competing interests}

The authors declare no competing interests.

\bmhead{Use of AI tools}

An AI assistant was used to help draft and edit the manuscript from author-supplied data and analysis. The authors are responsible for the final text, the analysis choices and the submission.

\bibliography{references}

\begin{appendices}
\renewcommand{\theHtable}{appendix.\thetable}

\section{Tasks}\label{app:tasks}

\begin{table}[htbp]
\caption{Tasks in the analysis set}\label{tab:tasks}
\small
\begin{tabular}{@{}lll@{}}
\toprule
Task & Domain & Difficulty \\
\midrule
3d-scan-calc & industrial and physical systems & hard \\
ada-bathroom-plan-repair & industrial and physical systems & hard \\
azure-bgp-oscillation-route-leak & software engineering & medium \\
bike-rebalance & mathematics and formal reasoning & medium \\
citation-check & office and white-collar work & medium \\
civ6-adjacency-optimizer & mathematics and formal reasoning & hard \\
crystallographic-wyckoff-position-analysis & natural science & medium \\
dapt-intrusion-detection & cybersecurity & hard \\
data-to-d3 & software engineering & medium \\
drone-planning-control & industrial and physical systems & medium \\
dynamic-object-aware-egomotion & industrial and physical systems & medium \\
earthquake-phase-association & natural science & hard \\
earthquake-plate-calculation & natural science & medium \\
econ-detrending-correlation & finance and economics & medium \\
edit-pdf & office and white-collar work & medium \\
energy-ac-optimal-power-flow & industrial and physical systems & medium \\
exam-block-sequencing & mathematics and formal reasoning & hard \\
exoplanet-detection-period & natural science & medium \\
fix-build-agentops & software engineering & easy \\
fix-build-google-auto & software engineering & easy \\
fix-druid-loophole-cve & cybersecurity & hard \\
fix-visual-stability & software engineering & hard \\
glm-lake-mendota & natural science & hard \\
invoice-fraud-detection & finance and economics & hard \\
jpg-ocr-stat & office and white-collar work & hard \\
mario-coin-counting & media and content production & medium \\
multilingual-video-dubbing & media and content production & medium \\
offer-letter-generator & office and white-collar work & easy \\
pddl-airport-planning & mathematics and formal reasoning & medium \\
pddl-tpp-planning & mathematics and formal reasoning & medium \\
powerlifting-coef-calc & office and white-collar work & easy \\
reserves-at-risk-calc & finance and economics & medium \\
sec-financial-report & finance and economics & hard \\
shock-analysis-demand & finance and economics & medium \\
software-dependency-audit & cybersecurity & medium \\
spring-boot-jakarta-migration & software engineering & hard \\
suricata-custom-exfil & cybersecurity & medium \\
syzkaller-ppdev-syzlang & cybersecurity & medium \\
threejs-structure-parser & media and content production & medium \\
video-silence-remover & media and content production & hard \\
\botrule
\end{tabular}
\end{table}

\end{appendices}

\end{document}